# 環境適応処理における配置再構成の提案

山登庸次†

† NTT ネットワークサービスシステム研究所，東京都武蔵野市緑町 3-9-11
E-mail: †yoji.yamato.wa@hco.ntt.co.jp

**あらまし**　ヘテロハードウェアを活用するには，プログラマーは OpenMP，CUDA，OpenCL 等の十分な技術スキルが必要であった．そこで，私は，一度記述したコードの自動変換，構成を行い，高性能操作を可能にする環境適応ソフトウェアを提案し，自動変換，適正配置等に取り組んできた．しかし，今まで変換アプリケーションをどこに初期配置するかは検討されていたが，他ユーザの配置状況を考慮した全体最適な配置は検討されていなかった．本論文では，環境適応ソフトウェアの新しい要素として，運用開始後に他ユーザの配置状況を勘案して全体的なユーザ満足度を向上させる運用中配置再構成を，線形計画手法を用いて検討する．再構成アプリケーション数等の条件を変化させたシミュレーション実験を通じて適切に再配置できることを確認した．

**キーワード**　環境適応ソフトウェア, 自動オフロード，最適配置，運用中再構成

# Proposal of deployment reconfiguration for environment adaptation

Yoji YAMATO†

† Network Service Systems Laboratories, NTT Corporation, 3-9-11, Midori-cho, Musashino-shi, Tokyo
E-mail: †yoji.yamato.wa@hco.ntt.co.jp

**Abstract**　To use heterogeneous hardware, programmers needed sufficient technical skills such as OpenMP, CUDA, and OpenCL. Therefore, I have proposed environment-adaptive software that enables high-performance operation by automatically converting and configuring the code once written, and have been working on automatic conversion and proper placement. However, until now, where to initially place the converted application has been considered, but the overall optimal placement has not been considered in consideration of the placement status of other users. In this paper, as a new element of environment-adaptive software, I study the relocation during operation, which improves the overall user satisfaction by considering the placement of other users, using a linear programming method. It was confirmed that it can be properly rearranged through simulation experiments.

**Key words**　Environment Adaptive Software, Automatic Offloading, Optimum Placement, Reconfiguration During Operation

## 1. はじめに

近年，ムーアの法則の鈍化が予想されている．そのような状況から，CPU だけでなく，FPGA（Field Programmable Gate Array）や GPU（Graphics Processing Unit）等のデバイスの活用が増えている．例えば，Microsoft 社は FPGA を使って Bing の検索効率を高め [1]，Amazon 社は，FPGA, GPU をクラウド技術を用いて（例えば, [2]- [8]）インスタンス提供している [9]．また，システムでの IoT 機器利用（例えば, [10]- [18]）も増えている．

しかし，CPU 以外のデバイスを適切に活用するためには，デバイス特性を意識したプログラム作成が必要で，OpenMP（Open Multi-Processing）[19]，OpenCL（Open Computing Language）[20]，CUDA（Compute Unified Device Architecture）[21] や組み込み技術といった知識が必要になってくるため，スキルの壁が高い．Java [22] は一度記述したコードをどこでも動かせると言うが性能は考慮外だった．そこで，私は，一度記述したコードを, 配置先の環境に存在する GPU や FPGA, メニーコア CPU 等を利用できるように，変換，配置等を自動で行い，アプリ（アプリケーション）を高性能に動作させることを目的とした，環境適応ソフトウェアを提案した．合わせて，環境適応ソフトウェアの要素として，アプリのループ文及び機能ブロックを，FPGA，GPU に自動オフロードする方式を提案評価している [23]- [28]．さらに，変換アプリをユーザの応答時間や価格要求を満たして，配置先を適正化する手法についても提案してきた．

本稿では，環境適応ソフトウェアの新たな要素として，変換して配置し動作させているアプリを，別ユーザの配置状況等を踏まえて再配置することで，全体的なユーザ満足度を上げる運用中再構成の手法を提案し，評価する．

## 2. 既 存 技 術

### 2.1 市 中 技 術

GPU の並列計算能力を一般用途にも使う GPGPU（General Purpose GPU）（例えば，[29] 等）を行うための環境として CUDA が普及している．CUDA は GPGPU 向けの NVIDIA 社の環境だが，FPGA，メニーコア CPU，GPU 等のヘテロなデバイスを同じように扱うための仕様として OpenCL とその開発環境（例えば，[30] 等）が出てきている．しかし，CUDA，OpenCL は，C 言語の拡張を行いプログラムを行う形だが，メモリ処理の記述等プログラムの難度は高い．

このため，スキルが乏しいプログラマーが，FPGA や GPU を活用してアプリを高速化することは難しいし，自動並列化技術（例えば，[31] 等）を使う場合も並列処理箇所探索の試行錯誤等の稼働が必要だった．並列処理箇所探索の試行錯誤を指示行ベース仕様（OpenACC [32] [33] 等）を用いて自動化する取り組みとして，著者は進化計算手法を用いた GPU 自動オフロードを提案している．

配置に関して，ネットワークリソースの最適利用として，ネットワーク上にあるサーバ 群に対して VN（Virtual Network）の埋め込み位置を最適化する研究がある [34]．[34] では，通信トラヒックを考慮した VN の最適配置を決定する．しかし，単一リソースの仮想ネットワークが対象で，キャリアの設備コストや全体的応答時間の削減が目的で，個々に異なるアプリの処理時間や，個々のユーザの価格や応答時間要求等の条件は考慮されていなかった．

### 2.2 環境適応処理のフロー

ソフトウェアの環境適応を実現するため，著者は図 1 の処理フローを提案している．環境適応ソフトウェアは，環境適応機能を中心に，検証環境，商用環境，テストケース DB，コードパターン DB，設備リソース DB の機能群が連携することで動作する．

Step1 コード分析：

Step2 オフロード可能部抽出：

Step3 適切なオフロード部探索：

Step4 リソース量調整：

Step5 配置場所調整：

Step6 実行ファイル配置と動作検証：

Step7 運用中再構成：

ここで，Step 1-7 で，環境適応するために必要となる，コードの変換，リソース量の調整，配置場所の決定，検証，運用中の再構成を行うことできる．

現状を整理する．ヘテロなデバイスに対するオフロードは手動での取組みが主流である．著者は環境適応ソフトウェアのコンセプトを提案し，自動オフロードを検討してきたが，自動変

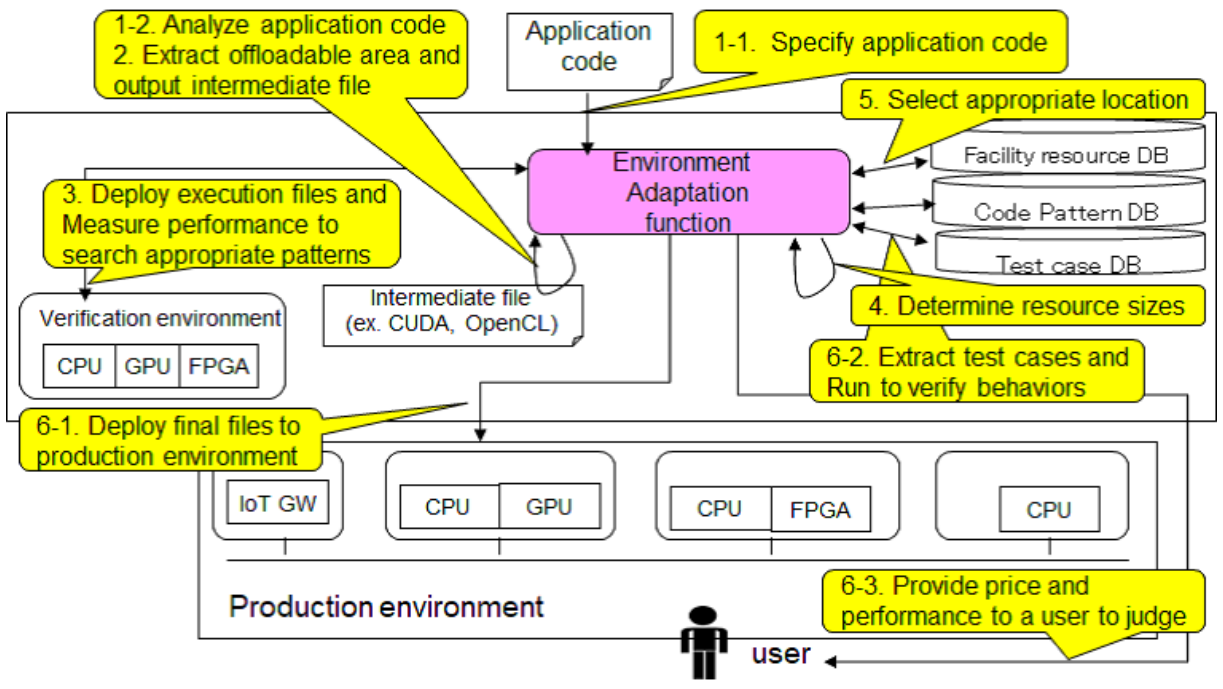

図 1　環境適応ソフトウェアのフロー

換した後のアプリの配置については，個別ユーザの要件に応じた個別最適の配置検討のみで，他ユーザの配置状況を勘案した全体最適の配置検討はされていない．そのため，本稿では，運用開始後に配置を再構成することで再構成対象の複数のユーザの満足度を向上させることを検討する．

## 3. アプリケーション配置再構成

著者は，環境適応ソフトウェアのコンセプトを具体化するために，これまで，プログラムループ文や機能ブロックの GPU，FPGA 自動オフロード，オフロード先リソース量自動調整や配置適正化を提案してきた．これらの要素技術検討も踏まえて，本節の，3.1 では，自動オフロードの例としてループ文の GPU 自動変換技術について概説する，3.2 では，既存の初期配置方式と再構成の必要性を検討する．3.3 では，アプリを適切に再配置するための，線形計画手法の定式化と方式提案を行う．

### 3.1 ループ文の GPU 自動変換

通常の CPU で動作しているプログラムを，GPU や FPGA 等のデバイスにオフロードすることで高速化する事例は多いが，自動で行っている例は少ない．ループ文の GPU 自動変換技術を紹介する．

まず，基本的な課題として，コンパイラがこのループ文は GPU で並列処理できないという制限を見つけることは可能だが，このループ文は GPU の並列処理に適しているという適合性を見つけることは難しいのが現状である．一般的にループ回数が多い等のループの方が適していると言われるが，実際に GPU にオフロードすることでどの程度の性能になるかは，実測してみないと予測は困難である．そのため，このループを GPU にオフロードするという指示を手動で行い，性能測定を試行錯誤することが頻繁に行われている．[23] はそれを踏まえ，Clang [35] 等でのパースライブラリで解析されたコード構造から，GPU にオフロードする適切なループ文の発見を，進化計算手法の一つである遺伝的アルゴリズム GA（Genetic Algorithm）[36] で自動的に行うことを提案している．GPU 処理を想定していない通常 CPU 向け汎用プログラムから，最初に並列可能ループ文のチェックを行い，次に並列可能ループ文群に対して，GPU 実行の際を 1，CPU 実行の際を 0 と値を置いて遺伝子化し，検証環境で性能検証試行を反復し適切な領域を探索している．並列可能ループ文に絞った上で，遺伝子の部

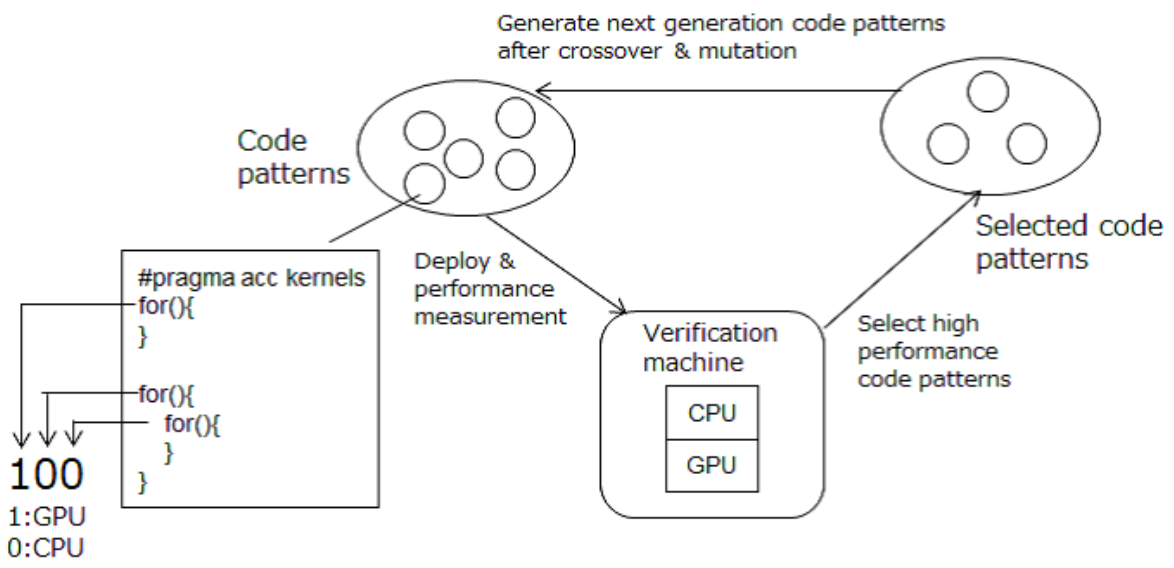

図 2　ループ文の GPU 自動変換方式

分の形で，高速化可能な並列処理パターンを保持し組み換えていくことで，取り得る膨大な並列処理パターンから，効率的に高速化可能なパターンを探索している（図 2）.

[28] では，ループ文の適切な抽出に加えて，ネストループ文の中で利用される変数について，ループ文を GPU にオフロードする際に，ネストの下位で CPU-GPU 転送が行われると下位のループの度に転送が行われ効率的でないため，上位で CPU-GPU 転送が行われても問題ない変数については，上位でまとめて転送を行うことを提案している.

更に，より CPU-GPU 転送を削減する方式も提案している．具体的には，ネストだけでなく，複数ファイルで定義された変数について，GPU 処理と CPU 処理が入れ子にならず，CPU 処理と GPU 処理が分けられる変数については，一括化して転送する指定を行う．さらに，コンパイラがコンパイラ判断で変数を転送してしまうことを避けるため，変数転送には一時領域を作成し一時領域でパラメータを初期化して，CPU-GPU 転送に用いている.

ループ文の GPU 自動変換については，進化計算手法を用いた最適化と，CPU-GPU 転送の低減により，自動化を可能としている.

同様の手法を用いて，ループ文の FPGA 向け自動変換，機能ブロックの自動オフロードを提案しており，自動変換後，コストパフォーマンス高く運用するためオフロード先デバイスのリソース量を自動調整する方式も提案している.

### 3.2　既存方式と再構成必要性

計算ノードリンクとしては，ユーザエッジ，キャリアエッジ，クラウドの 3 層のトポロジーを想定する（図 3 等）．計算ノードは CPU，GPU，FPGA の 3 種に分けられる．GPU や FPGA を備えるノードは，仮想化技術（例えば，[37]）により，GPU，FPGA インスタンスとして，CPU リソースも含む形で提供される.

従来は [34] 等の検討の様に，例えば仮想ネットワークを収容するサーバを配置する場所を，トラフィック増加等の長期的傾向を見て，計画的に設計していた．それに対して，環境適応ソフトウェアでは 2 つ特徴がある．一つ目は，配置アプリは静的に定まっているのでなく，GPU や FPGA 向けに自動変換され，GA 等を通じ利用形態に適したパターンが実測を通じて抽出されるため，アプリコードや性能は動的に変わり得る．二つ目は，キャリア設備コストや全体的応答時間だけ低減すれば良いのでなく，応答時間や価格に対する個々のユーザ要求を満たす必要があり，アプリ配置ポリシーも動的に変わり得る.

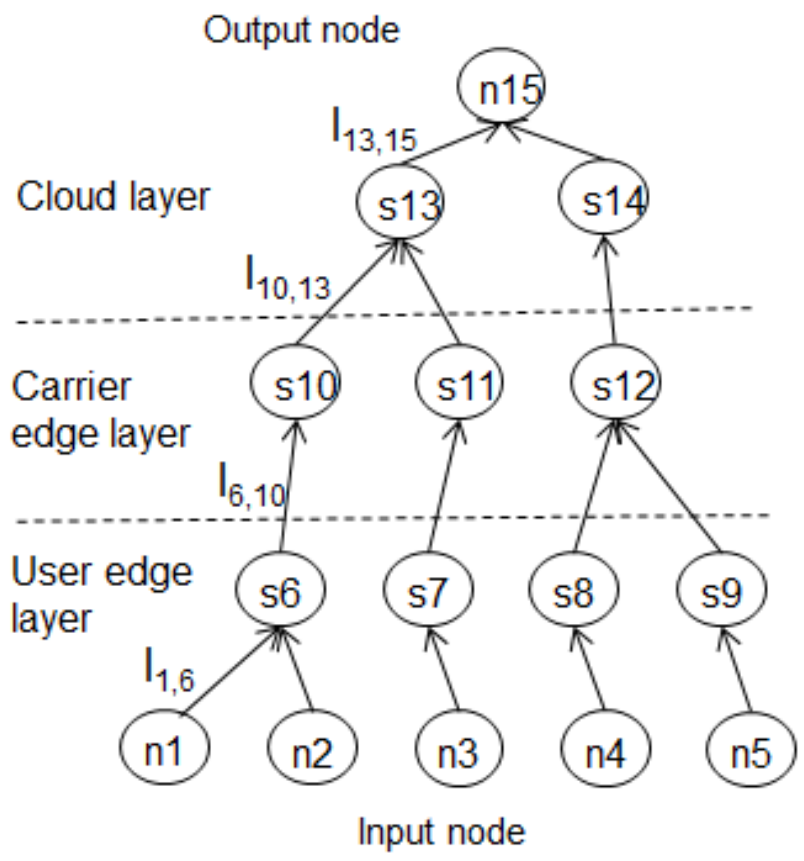

図 3　計算ノードのトポロジー例

この二つの特徴も踏まえ，著者は以前に，ユーザの応答時間要求や価格要求を満たす配置計算のため，線形計画手法により，ユーザがアプリを配置依頼した際の，変換後アプリの応答時間と価格を定式化して，応答時間か価格の片方を目的関数に最小化する方法を提案している．提案方式を用いて，複数種のアプリを想定して，1000 アプリの配置最適化を GLPK（Gnu Linear Programming Kit）5.0 にてシミュレーションし，提案方式の有効性を確認している.

しかし，提案方式は，個別ユーザの応答時間や価格要求に従い，アプリを配置していくため，基本的に早い者勝ちとなり，例えば安さ優先の要求条件ばかりの場合クラウドが，速さ優先の要求条件ばかりの場合エッジが埋まってしまい，埋まった後は別サーバへの配置が必要になってくる．そこで，早い者勝ちの配置を緩和するため，運用開始前だけでなく運用開始後の配置再構成が必要と考える.

### 3.3　再構成方式

本サブ節では，他ユーザ配置状況も踏まえ，アプリを適切な配置場所に再構成するための再構成方式を，線形計画手法の定式化も含めて提案する.

再構成では，一定数アプリの配置毎（100 アプリ等）に，複数ユーザの当初要求条件を満たす形で，一定数アプリ（全アプリ or100 アプリ等）の配置再構成を試行計算することで，応答時間や価格の変化により定まるユーザ群の満足度を向上させる．再構成の試行計算の結果，満足度変化が一定の閾値を超える等，再構成の効果が高い場合のみ，再構成が行われる．実際の再構成は，アプリ実行サーバの変更が必要となるため，ライブマイグレーション等の手法を用いて，ユーザ影響を抑えて行う.

再構成のための線形計画式を (1)-(5) に，パラメータを図 4 に示す．(1) がユーザ満足度評価のための目的関数，(2)(3) がアプリ配置毎にユーザが指定する応答時間要求，価格要求の制約条件，(4)(5) がサーバリソース上限の制約条件である.

まず，新規配置は (2)-(5) の式に応じて配置がされる．ユーザは配置の際に，応答時間，価格の片方か両方の要求条件を付

$a_i$: Device usage cost
$b_j$: Link usage cost
$C_i^d$: Device calculation resource limit of #i
$C_j^l$: Link bandwidth limit of #j
$C_k$: Data size of #k application
$A_{i,k}^d$: Whether to use of #k application on #i device
$A_{j,k}^l$: Whether to use of #k application on #j link
$B_k^d$: Calculation resource of #k application
$B_k^l$: Bandwidth usage of #k application
$B_{i,k}^p$: Processing time of #k application on #i device

図 4　線形計画式パラメータ

けることができる．応答時間要求 $R_k^{upper}$ は何秒以内に応答が必要か，価格要求 $P_k^{upper}$ は一月幾ら以下かを指定する．(2)(3) の片方だけ指定の場合は，その逆の最小化が目的関数となり，両方指定の場合は目的関数はどちらかの最小化をユーザが指定する．(4)(5) は，計算リソース及び通信帯域の上限を設定する制約条件であり，他ユーザ配置アプリ含めて計算され，新規ユーザのアプリ配置によるリソース上限超過を防ぐ．新規配置は，ユーザの配置依頼に対して，順次 (2)-(5) の計算を行っていくことで行われる．

次に再構成を考える．再構成は (1)-(5) 式に応じて計算がされるが，特にユーザ満足度に応じる値 S の計算が新たに加わる．個別ユーザの満足度として，再構成前の配置で応答時間 1 点，価格 1 点を基本の値として，再構成前応答時間 $R_k^{before}$ が再構成後 $R_k^{after}$ で X 倍になったら X が応答時間満足度に関連した値，再構成前価格 $P_k^{before}$ が再構成後 $P_k^{after}$ で Y 倍になったら Y が価格満足度に関連した値とする．再構成の試行計算の目的関数は，再構成対象の一定数アプリのユーザ群満足度に関連した値であり，複数アプリに対して（X+Y）の総和を最小化する配置を計算する．目的関数の具体的内容は (1) である．また，新規配置時に，ユーザが片方しか制約条件を指定していない場合は，そのアプリでは (2)(3) は片方だけ指定される．

(1)-(5) の線形計画の式に基づき，GLPK や CPLEX（IBM Decision Optimization）等の線形計画ソルバで解を導出することで，再構成の効果を計算する．再構成対象のアプリ数は一定値であり，全アプリでない事もある．ソルバ計算時間は，再構成を計算するアプリの数が増えると増大する．そのため，アプリ一定数の設定は可変とし，100 配置毎に 100 アプリを最適化や，一度に全アプリを最適化等は，ソルバの計算時間に応じてサイズを調整して決定する．

$$S = \sum_{k \in App} \left( \frac{R_k^{after}}{R_k^{before}} + \frac{P_k^{after}}{P_k^{before}} \right) \tag{1}$$

$$\sum_{i \in Device} (A_{i,k}^d \cdot B_{i,k}^p) + \sum_{j \in Link} (A_{j,k}^l \cdot \frac{C_k}{B_k^l}) = R_k^{after} \leqq R_k^{upper} \tag{2}$$

$$\sum_{i \in Device} a_i \left( \frac{A_{i,k}^d \cdot B_k^d}{C_i^d} \right) + \sum_{j \in Link} b_j \left( \frac{A_{j,k}^l \cdot B_k^l}{C_j^l} \right) = P_k^{after} \leqq P_k^{upper} \tag{3}$$

$$\sum_{k \in App} (A_{i,k}^d \cdot B_k^d) \leqq C_i^d \tag{4}$$

$$\sum_{k \in App} (A_{j,k}^l \cdot B_k^l) \leqq C_j^l \tag{5}$$

# 4. 評　　価

## 4.1 評価条件

### 4.1.1 対象アプリケーション

配置アプリは，多くのユーザが利用すると想定されるフーリエ変換と画像処理とする．

フーリエ変換処理 FFT（Fast Fourier Transform）は，振動周波数の分析等，IoT でのモニタリングの様々な場面で利用されている．NAS.FT [38] は，FFT 処理のオープンソースアプリの一つである．備え付けのサンプルテストの 2048*2048 サイズの計算を行う．

MRI-Q [39] は，非デカルト空間の 3 次元 MRI 再構成アルゴリズムで使用されるキャリブレーション用のスキャナー構成を表す行列 Q を計算する．MRI-Q は，パフォーマンス測定中に 3 次元 MRI 画像処理を実行し，Large の 64*64*64 サイズのサンプルデータを使用して処理時間を測定する．

著者の GPU，FPGA 自動オフロード技術 [28] により，NAS.FT は GPU で，MRI-Q は FPGA で高速化でき，CPU に比べてそれぞれ 5 倍，7 倍の性能がでることが分かっている．

### 4.1.2 評価手法

アプリを配置するトポロジーは図 3 の様に 3 層で構成され，クラウドレイヤーの拠点数は 5，キャリアエッジレイヤーは 20，ユーザエッジレイヤーは 60，インプットノードは 300 とする．

各拠点に，クラウドでは，サーバは CPU 8 台，GPU 16GB RAM 4 台，FPGA 2 台，キャリアエッジでは，CPU 4 台，GPU 8GB RAM 2 台，FPGA 1 台，ユーザエッジでは，CPU 2 台，GPU 4GB RAM 1 台とする．サーバ 1 台の全リソース使用（GPU では 16GB RAM 利用時）の月額はクラウドでは 5 万，10 万，12 万とした．集約効果のため，キャリアエッジ，ユーザエッジは割高になり，クラウドの 1.25 倍，1.5 倍の月額とした．

リンクについては，クラウド-キャリアエッジ間は 100Mbps，キャリアエッジ-ユーザエッジ間は 10Mbps の帯域が確保されている．リンクコストについては，100Mbps のリンクは月額 8000 円，10Mbps のリンクは月額 3000 円とした．

アプリが利用するリソースとして，処理時間等は，実際に GPU，FPGA にオフロードした際の値を用いる．NAS.FT は，利用リソース量は GPU 1GB RAM，利用帯域 2Mbps，転送データ量 0.2MB，処理時間 5.8 秒である．MRI-Q は，利用リソース量は FPGA サーバの 10%（FlipFlop，LookUpTable 等の利用数が FPGA の利用リソースとなる），利用帯域 1Mbps，転送データ量 0.15MB，処理時間 2.0 秒である．

まず，400 個のアプリを新規配置する．アプリは，インプットノードから生じるデータを分析する想定で，300 のインプットノードから配置依頼をランダムに生じさせる．配置依頼数として，NAS.FT：MRI-Q=3:1 の割合で 400 回アプリの配置依頼をする．

ユーザ要求条件として，配置依頼する際に価格条件か応答時間条件かその両方が選ばれる．NAS.FT の場合，価格に関しては月 7500 円 (a) か 8500 円 (b) か 10000 円 (c) 上限か，応答時間に関しては 6 秒 (A) か 7 秒 (B) か 10 秒 (C) 上限かが選択される．MRI-Q の場合，価格に関しては月 12500 円 (x) か 20000 円 (y) 上限か，応答時間に関しては，4 秒 (X) か 8 秒 (Y) 上限が選択される．ユーザ要求として，NAS.FT では a, b, c, A, B, C, aC, bB, bC, cA, cB, cC をそれぞれ 1/12 ずつ，MRI-Q では x, y, X, Y, xY, yX, yY をそれぞれ 1/7 ずつの確率で選択する．ユーザ上限要求が 1 つの指標の際は別指標の最小化が目的関数に，2 つの指標の際は一方がランダムに選択されその最小化が目的関数になる．

次に，400 個の配置後は，100 アプリ配置毎に，一定数アプリの再構成を行う．配置周期の配置アプリ数は 100 で固定だが再構成アプリ数は，100，200，400 アプリと可変にし，ユーザ満足度を計算する．

### 4.2 結　　果

実験は，ソルバ GLPK5.0 を用いたシミュレーションにより行った．図 5(a) は，再構成対象アプリ数を横軸に，実際に再構成したアプリ数を縦軸に取ったグラフであり，図 5(b) は，実際に再構成したアプリの $R_k^{after}/R_k^{before}+P_k^{after}/P_k^{before}$ の平均値を縦軸に取ったグラフである．

計算時間について，新規配置では総計 500 個を順に計算して配置するだけなので，1 分以内の短時間で終わる．一方，再構成時間は，再構成対象のアプリが増えるにつれ，線形計画の条件式が増えることになるが，100 アプリでは 10 秒以内だったが，400 アプリでも 1 分以内で終わっている．

図 5(a) に関して，若干ばらつきはあるが，再構成対象アプリ数の約 1 割が，実際に再構成されていることが分かる．新規配置では，1 アプリ毎に個別最適な配置をしているが，再構成では複数アプリ一括で適正配置を計算することで，再構成可能なアプリがある程度見つかることが分かる．図 5(b) に関して，再構成を行ったアプリは $R_k^{after}/R_k^{before}+P_k^{after}/P_k^{before}$ の平均が 1.96 程度に改善されていることが分かる．この値は 2 から大きく改善される値ではないが，例えば，NAS.FT をキャリアエッジからクラウドに配置を変えた際は応答時間が 6.6 秒から 7.4 秒になるが，価格が約 8400 円から約 7000 円となるため，値は 2 から 1.954 になる．図 5(b) より，この値は再構成対象アプリ数によらずほぼ一定となっており，再構成対象は全アプリを対象にしなくても良いことが分かる．

## 5. ま と め

本稿では，著者が提案している環境適応ソフトウェアの新たな要素として，運用開始後に他ユーザのアプリ配置状況を勘案して，再配置を行うことで，対象ユーザの満足度を向上させる

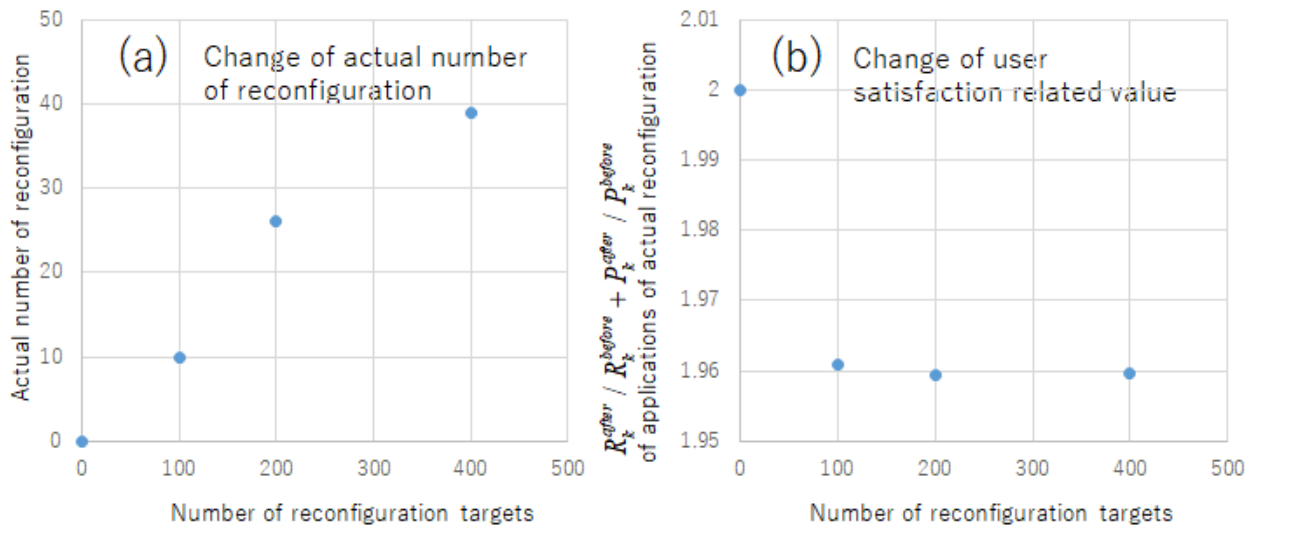

図 5 (a) 実再構成アプリ数の変化，(b) 実再構成アプリの $R_k^{after}/R_k^{before}+P_k^{after}/P_k^{before}$ の変化

再構成方式を提案した．

提案方式は，線形計画手法を用いて，再構成対象アプリのユーザ群満足度の向上を目的関数に，試行計算する．具体的には，ユーザの応答時間，価格の要求条件を満たした上で，再構成した際の応答時間，価格から定まるユーザ満足度を計算し，線形計画ソルバで再構成対象の複数アプリに対する解を求める．

複数種のアプリを想定して，シミュレーション実験を行い，提案方式で再構成した際のユーザ満足度向上を確認し，提案方式の有効性を示した．今後は，配置だけでなく，オフロードデバイスのリソースバランス再構成等，再構成対象の拡張を検討する．

## 文　　献